\documentclass[conference]{IEEEtran}
\usepackage{cite}
\usepackage{amsmath,amssymb,amsfonts}
\usepackage{algorithmic}
\usepackage{graphicx}
\usepackage{textcomp}
\usepackage{xcolor}
\def\BibTeX{{\rm B\kern-.05em{\sc i\kern-.025em b}\kern-.08em
T\kern-.1667em\lower.7ex\hbox{E}\kern-.125emX}}
\begin{document}

\title{Anytime Learning - The next Step in Organic Computing?\\}

\author{\IEEEauthorblockN{Lucas Breitsameter}
\IEEEauthorblockA{\textit{University of Passau}\\
Passau, Germany \\
breits06@gw.uni-passau.de }
}

\maketitle

\begin{abstract}
Anytime learning describes a relatively novel concept by which systems are able to acquire knowledge about a changing environment and adapt and behave accordingly to this. "Anytime" refers to the fact that the system is capable of returning imperfect results at any point in time, which allows it to remain functional even if a perfect solution could not be found within the necessary time frame. This paper focuses on illustrating the concept of anytime learning and examining how it relates to organic computing as a whole. Could anytime learning be the next step in organic computing? 
\end{abstract}


\section{Motivation}
It has long been established that technical systems are growing more complex at a rapid pace. In fact, some argue that the complexity is approaching a point at which it would be unfeasible or impossible to have systems be manually operated by humans, as it is done in our current time. One of the first major companies to come to this realization was IBM, back in 2001 [1]. Their findings by 2003 indicated that failing to take action to deal with this matter would lead to immense economic costs and claimed "[t]he spiraling cost of managing the increasing complexity of computing systems is becoming a significant inhibitor that threatens to undermine the future growth and societal benefits of information technology."[2] Fig. 1 was published in [2] to illustrate the amount of costs that are closely linked to the increase in complexity and could potentially be prevented by utilizing more efficient methods. The costs in the figure were sorted according to industry sectors.
In order to counteract the "looming complexity crisis" as it was referred to by some, IBM proposed the concept of autonomic computing in 2003 [1], according to which systems should become more life-like and less prone to errors or unexpected situations. 
Proposals for similar concepts, such as organic computing, soon followed. All of them had one common goal, which also was indicated by the biological connotation of the terms: Bring systems closer to living organisms. "The technology needs to manage itself. So instead of the technology behaving in its usual pedantic way and requiring a human being to do everything for it, it starts behaving more like the 'intelligent' computer we all expect it to be and starts taking care of its own needs", as Irving Wladawsky-Berger from IBM phrased it at a conference in 2001 [2].

Ambitious as this sounds, science is still in the midst of researching the matter and finding tangible solutions for a broad variety of systems which are affected by this issue. Although it took IBM and most other instances until the early 2000s to acknowledge these problems, it appears one approach may have been far ahead of its time and dealt with the matter quite a bit earlier than that. This promising approach was "anytime learning" and was proposed by John J. Grefenstette and Connie Loggia Ramsey [3] as early as 1992. In this paper, that particular approach will be examined in detail. The design, strengths and weaknesses of anytime learning will be examined to answer the question: Could anytime learning be the next step in organic computing?

\begin{figure}[htbp]
\graphicspath {./images/}
\centerline{\includegraphics[width=3.5in]{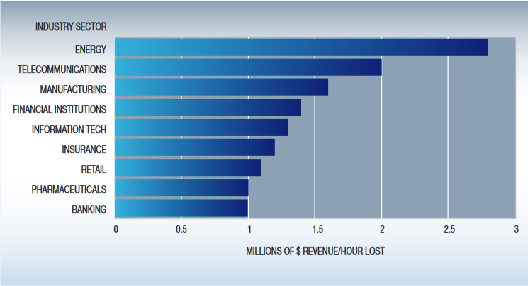}}
\caption{Costs which might be prevented with autonomic computing according to IBM, found in [2]}
\label{fig}
\end{figure}

\section{Origins of Anytime Learning}
While anytime learning itself is a relatively novel concept, its roots can be found much earlier. 
Some of the biggest contributors to the idea were Dean and Boddy, whose work in the area of anytime algorithms served as a basis for the concept of anytime learning [3]. 
Therefore, in order to understand the concept of anytime learning, it is essential to understand the concept of anytime algorithms. 
As is explained in [4], anytime algorithms can be utilized whenever it is not feasible or desirable to find the optimal solution to a problem. Their main characteristic is that they are able to return imperfect results if circumstances require for it, and the more time they are given to find a solution, the better the solution will become. This concept is particularly relevant to anytime learning, as we are dealing with an artificial intelligence which is very unlikely to ever return a perfect result within reasonable time and, as it will often operate in real-time, has a very limited amount of time to respond to external inputs. It therefore makes sense to construct a system which is capable of returning a reasonable response on demand at any point in time.

Other influences which are mentioned in Grefenstette's and Ramsey's work are iterative improvement techniques such as genetic algorithms or reinforcement methods. 

\section{Concept}
Since the concept of anytime learning is typically applied in conjunction with robots and artificial intelligence, a meaningful part of it revolves around hardware, such that the term "agent" may refer to any physical device, for instance a robot. 
The issue anytime learning deals with is systems having to continuously adapt to a changing environment while simultaneously remaining functional and being able to make reasonable decisions at any point in time. The components and methods,which make anytime learning achieve this, are explained in this section.

\subsection{Structure}

The basic idea, which allows for the concept of anytime learning to work, is the separation of the system into two core components communicating with each other. Those modules are the execution (or control) module, along with the hardware attached to an agent, as well as the learning module. Fig. 2 from [5] shows a simplified form of this described model.
\begin{figure}[htbp]
\centerline{\includegraphics[width=3.5in]{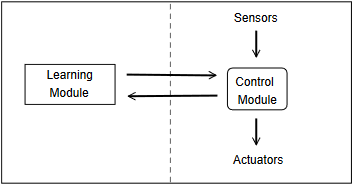}}
\caption{Simplified anytime learning architecture found in [5]}
\label{fig}
\end{figure}

For better understanding, the following description of the more detailed form of the model is visualized in Fig. 3, which was originally published in [3].

The execution system is responsible for \textit{monitoring} the \textit{environment}, \textit{making decisions} based on its findings, as well as providing the learning system with information regarding the current state of the environment. 

The learning system is, as the name implies, the component implementing the \textit{learning method} itself. Based on the environmental data supplied by the execution system, the \textit{simulation model} used by the learning algorithm is modified. Analogously, the execution system may then be modified based on the results of the learning process.

\subsection{Communication}
With the components in place, the next important characteristic about anytime learning is the way how those components communicate with each other. Grefenstette and Ramsey [3] describe two different types of communication. The first form of communication is initiated by the learning system and targets the execution system. This communication should occur whenever a better strategy for the agent was found, in order to keep the execution module updated on the best possible strategy. The reasoning for this is obvious: In case a better strategy is found, the execution system should obviously begin using the new strategy as soon as possible. 

Secondly, some communication is initiated by the execution system, or its monitor, to be more specific. A change of the environment which is detected by the monitor would be extremely relevant to the simulation, since the simulation model in turn is used by the learning algorithm. Therefore, if the monitor detects a meaningful change, this indicates that the simulation model must be adjusted and the monitor notifies the learning system of this fact, which will then respond with the appropriate action, like adjusting its simulation model or starting the learning process anew. 

\begin{figure}[htbp]
\centerline{\includegraphics[width=3.5in]{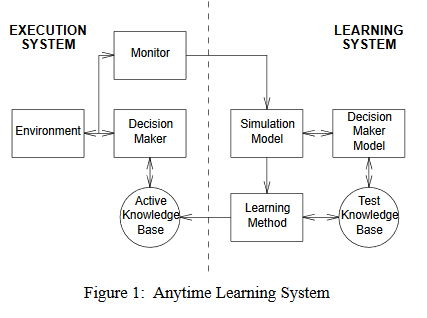}}
\caption{Detailed anytime learning architecture found in [3]}
\label{fig}
\end{figure}

\subsection{Policies}

Finally, according to [3] the system requires two particular policies to be complete. Firstly, it must be defined under which conditions the learning system has to adjust its simulation model. A drop-off in performance may indicate the environment has changed and the model should be adjusted, but that this is decision is correct can not be guaranteed. Similarly, improvements in performance are obviously intended and desired, but if those improvements occur unexpectedly, it is once again possible the environment changed, and changing up the model used for learning may result in even better results. 

The second policy must define how the learning system is adjusted, in case the first policy suggests it should. Which behavior is appropriate for this heavily depends on the internal workings of the learning system, which may vary for each implementation of anytime learning, such that a feasible response to this could be anything ranging from restarting the system to smoothly adjusting the model. 

\section{Application}
Grefenstette and Ramsey [3] tested the concept in a case study which they referred to as a "game of cat-and-mouse". It was organized as follows: A \textit{tracker} (cat) agent was given the task to track and follow a \textit{target} (mouse), but without ever getting closer than a certain threshold, or else the target would run away from the tracker at once with high speed.

For this study, they implemented anytime learning for the tracker and compared its performance to a tracker which was using the same learning algorithm, but operated without a monitor. The tracker without a monitor therefore disregarded its environment and effectively limited the learning algorithm to using a static model of the environment. This is the way most systems without anytime learning would behave like and is referred to as \textit{baseline learning}. \textit{Environment} in this case refers to the speed of the mouse, which, of course, is part of the environment that is under surveillance of the \textit{monitor} and may therefore be used for learning processes. The learning system was based on SAMUEL, which was developed by Grefenstette. The details on SAMUEL can be found in [6], but for this paper, it suffices to know that it is a program using genetic algorithms in order to learn and evolve. When anytime learning is implemented in a different context, SAMUEL may theoretically be substituted with any other learning method.

\subsection{Results}
The case study yielded the results shown in fig. 4.

\begin{figure}[htbp]
\centerline{\includegraphics[width=3.5in]{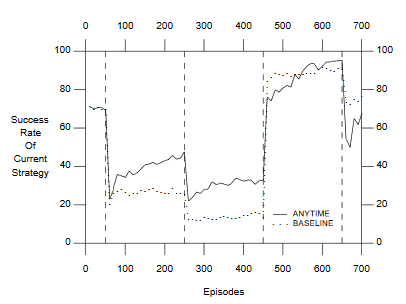}}
\caption{Results of the cat-and-mouse case study using basic anytime learning, published in [3]. The dashed vertical lines mark the points in time at which the environment was changed. "Episode" refers to a time frame during which the behavior of the agents was tested and evaluated.}
\label{fig}
\end{figure}

As can be seen, the agent using anytime learning generally performs better than the one that does not. The only exception to this occurs when the environment has only just changed, forcing the anytime learning agent to reset its learning progress since it is not applicable in the new environment. The agent which is not using anytime learning to begin with is not affected by this, as its learning has happened independently from its environment since the very start.

\section{Challenges}
While it has been shown that anytime learning can perform well under the right conditions, there are a number of challenges involved in implementing it efficiently.

For one, the exact implementation of the system is left up to the developer and depends heavily on the context in which it is being used. It can be assumed that finding the optimal implementation of components for a particular application is going to be rather difficult and therefore requires special attention.

Secondly, as was visible in the results of case study, in a rapidly changing environment, or even just an environment quickly alternating between two different states, anytime learning may fall short compared to other methods, as the learning progress is typically lost as soon as the environment changes.

And finally, as is pointed out in [3], it is crucial that the environment is parameterized accurately enough for the simulation to be able to adapt as is needed. This is also referred to as the "system identification problem" in other literature and as of now, no perfect solution to it has been found.

\section{Variants of Anytime Learning}
Based on Grefenstette and Ramsey's proposed approach, a number of different versions of anytime learning have been developed. Some notable ones and how they compare to the basic system described above will be briefly explained in this section. 

\subsection{Case-Based Anytime Learning}
\begin{figure}[htbp]
\centerline{\includegraphics[width=3.5in]{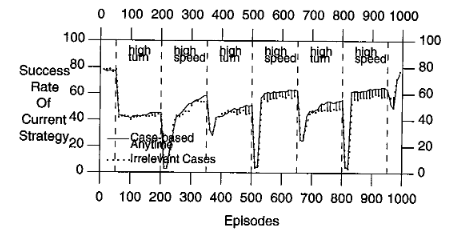}}
\caption{Results of the cat-and-mouse study using case-based anytime learning, published in [7]}
\label{fig}
\end{figure}

The idea behind the concept of case-based anytime learning [7] is rather simple. Instead of having to reset the learning process whenever the environment changes, a nearest neighbor algorithm is used in order to greatly increase the initial performance of the agent, if it is confronted with an environment which it previously encountered in a similar form. The results generated by this changed model in the cat-and-mouse study can be found in Fig. 5, which was published in [7].

What can be noticed is an increase in performance in a changing environment, compared to the basic anytime learning algorithm. This change allows it to keep up with other algorithms which do not depend on their environment while at the same time retaining all the advantages the original anytime learning model had. The biggest downside of this model is the immense amount of storage required to remember its previous learning and map them to an environment.

\subsection{Punctuated Anytime Learning}

In [5], Parker suggests an architecture for anytime learning which can be used by agents facing high risks. Rather than developing expensive agents which all require the components to implement anytime learning by themselves, he suggests an approach by which the learning system is ran by a central computer, while an observer periodically checks on the agents in order to update the central learning module. Fig. 6 from [5] shows a visualization of this concept.
The advantage is obvious: The agents become inexpensive while at the same time retaining their anytime learning capabilities, albeit to a lesser extent. While the system does not mimic a model in which each agent is equipped with its own learning module perfectly, it "can be an effective means of coupling the learning system to the robot during evolutionary computation" according to Parker [5].

\begin{figure}[htbp]
\centerline{\includegraphics[width=3.5in]{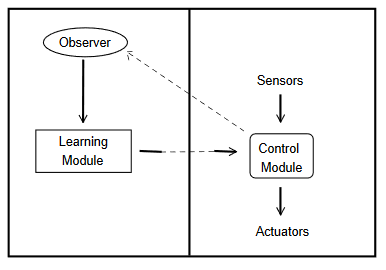}}
\caption{Punctuated Anytime Learning, found in [6]}
\label{fig}
\end{figure}

\subsection{Anytime Learning and adaptation of structured fuzzy behaviors}
In another architecture [8], anytime learning in conjunction with structured fuzzy behaviors was used as a means of successfully increasing the performance of a reinforcement learning algorithm being used by robots, which otherwise had not been feasible due to performance problems caused by the typically slow performance of reinforcement learning. This indicates that a lot of algorithms which already exist could be greatly improved by anytime learning.

\subsection{Summary}
Although these are just a few examples of how anytime learning can be utilized and improved, it goes to show how flexible the concept itself is and how it can be applied to a large variety of situations in order to deal with a number of issues which previously posed a problem.

\section{Relation to Organic Computing}
\subsection{Introduction to Organic Computing}
In order to understand which meaning the concept of anytime learning holds to the world of organic computing as a whole, it is of course necessary to define what the term organic computing entails to begin with. This task is more challenging than one might assume, since it is a very broad and relatively novel topic in computer science. The meaning of the term "Computing" is rather obvious, whereas "Organic" requires some explanation. In general, Organic is to be understood as "life-like". That is to say: Organic systems behave akin to biological organisms, which are defined by their ability to dynamically adapt to changes and therefore have an extremely high tolerance for error. Put very simply: If an organic organism, such as an animal, fails to perform a task, it will not suddenly drop dead on the spot because its system ended up in an unexpected state. Much rather, it would try to correct the error which occurred, adjust accordingly and ideally even prevent the error from happening a second time. Since this works so well in nature, it is desirable to equip technical systems with very similar properties. It is the goal of organic computing to research these properties, apply them to technology, and ultimately, make technical systems more life-like this way.

\subsection{Self-x properties}
Organic and autonomic  systems are closely linked. According to [1], the properties an autonomic system should exhibit for it to be considered as such, are the so called self-x properties. In [9], Branke et al. explain how very similar properties apply to organic computing. They claim that "AC [Autonomic Computing] pursues very similar goals, but so far focuses on server architectures as application, while OC [organic computing] focuses on distributed, self-organizing technical systems." 
Since anytime learning clearly focuses on technical systems rather than server architectures, it would therefore be considered organic computing if the self-x properties apply to a reasonable degree. The amount of self-x properties that could be examined is immense and as the topic is still being researched, more are being added continuously. This is why for the purposes of this paper, only some of the most prominent self-x properties will be considered. Among these properties, according to [1], [9] and [10], are the following: self-optimization, self-configuration, self-healing, self-protection. Furthermore, although it is not strictly a self-x property, context-awareness is named by Mueller-Schloer [10] as another vital property of organic computing. The following section will discuss the meaning of these properties and in which way they apply to anytime learning.

\subsubsection{Self-optimization}

\textit{Self-optimizing} refers to the fact that a system should at all times be optimizing itself with as little manual input as possible, even after it has been deployed and is being used [1]. Typically this is accomplished by equipping systems with sophisticated learning mechanisms which allow them to learn from their past mistakes. 
In the case of anytime learning, these learning mechanisms can be found in the learning module, as its name suggests. As the system spends time interacting with its environment, it will improve its behavior according to the learning module with the clear goal to optimize itself. It is therefore safe to assume that anytime learning can be considered to be self-optimizing.

\subsubsection{Self-configuration}

\textit {Self-configuring} systems are supposed to be configured by following high-level policies and adjust the rest of the system automatically [1]. Rather than configuring each component individually, the system is capable of growing \textit{organically}. Components can be exchanged, added or removed at will, but throughout all this, the system as a whole remains functional and performs any additional required adjustments automatically.
Whether this property is being realized by anytime learning may be viewed with skepticism. It is true that a system using anytime learning will operate using a dynamic simulation model and continuously adjusts the parameters on this model in accordance to the environment, which is clearly a behavior that would be considered self-configuring. However, something else must be considered as well: anytime learning introduces a number of components to the system which heavily rely on one another, thereby increasing the difficulty of  modifying single components. One could argue that by doing so, the amount of self-configuration in the system is thereby even being decreased. Naturally, these problems may or may not occur, depending on which concepts are being used in addition to anytime learning. Nevertheless, at the end of the day, anytime learning itself does not suggest any mechanisms in particular to avoid these problems. To summarize: anytime learning can potentially enhance the self-configuring properties of a system slightly, but it does not appear to be the main focus of the concept.

\subsubsection{Self-healing}

\textit{Self-healing} systems aim to be highly resistant to errors in both hard- and software and are supposed to find and fix any problems in the system by themselves [1].
Anytime learning may not appear to have this property at first glance, but if the matter is examined further, it quickly becomes evident that this is one of the core objectives of anytime learning. As has been highlighted in [5], it seems likely that anytime learning could find application in agents which operate in dangerous environments. With the help of anytime learning, a system which detects that it would perform poorly in its active environment by retaining its current behavior, will change accordingly. In other words: The system recognizes the flaws in the model which it is using and corrects, ergo \textit{heals} them. In some cases, this could prevent serious damage to the hardware or the system as a whole. All of this leads to the conclusion that even though it is not immediately noticeable, anytime learning contributes immensely to the self-healing properties of a system. 

\subsubsection{Self-protection}

\textit{Self-protecting} systems are able to defend themselves against malicious attacks or erroneous user input. Furthermore, they will also "anticipate problems based on early reports from sensors and take steps to avoid or mitigate them." [8] 
When looking at anytime learning, it quickly becomes apparent that similar reasons which make anytime learning self-healing make it \textit{self-protecting} as well. The system protects itself against dangerous environments by adapting to them. All of this happens based on early findings of its sensors, which matches the above definition of self-protecting systems quite precisely. Furthermore, the anytime approach used by the system is a strong indicator of it being self-protecting. The advantage of anytime algorithms lies in the fact that they can return a valid result, even if they are interrupted before finishing. In other systems, this might lead to unpredictable and unexpected behavior, but with anytime learning, this is exactly what is being accounted for, making it very resilient to being interrupted, and therefore self-protecting.

\subsubsection{Context-awareness}

\textit{Context-awareness} plays a major role in organic computing [10]. A number of self-x properties rely on the system being able to monitor its surroundings and act accordingly. As such, context-awareness or \textit{monitoring} is often considered to be "an essential feature of autonomic elements" [1] and is a recurring feature found in numerous technologies related to organic computing, with anytime learning being among them. Of all properties, this is perhaps the one which defines anytime learning like no other. The very core idea of the concept is to monitor the environment and thus become aware of the context which the system is operating in. This knowledge is then used to enforce the self-x properties. As such, it is extremely obvious that anytime learning allows for a high degree of context-awareness and it seems likely that it was designed with specifically this purpose in mind. 

In summary, it is evident that anytime learning complies with the vast majority of these required properties. Even though some properties are more prevalent than others, as a whole it exhibits the characteristics which are so relevant for organic computing to a high extent. According to the self-x properties, the concept of anytime learning can therefore be classified as organic computing.

\subsection{Similarities to other architectures}
Apart from the self-x properties, one may also consider similarities to other architectures when discussing to which extent anytime learning could be considered organic computing. Once more, the number of architectures that could be examined is immense, which is why this paper will focus on just one of them: The MLOC model as it is suggested by Mueller-Schloer and Tomforde [11]. 

The MLOC (Multi-Level Organic Computing) architecture consists, as its name implies, of three connected layers, which can be seen in Fig. 7 [11]. These layers are the following: The reflective layer, the reactive layer and the active layer. MLOC is considered to be an architectural template for organic computing and as such, comparing it to the architecture of anytime learning yields some interesting results. What follows below is a more detailed examination of the single layers of MLOC and how anytime learning relates to them.

\subsubsection{Active layer}
On its first layer, the \textit{active} layer, the MLOC architecture consists of a System under Observation and Control (SuOC). Generally, the SuOC is a system which is influenced by its sensors and, by extent, the environment which it operates in. It is obvious how this closely resembles the way in which anytime learning works. Although there is no component which directly maps to this, the union of \textit{environment}, \textit{monitor} and the resulting data collected from the environment could be considered to be the equivalent of a SuOC in the context of anytime learning.

\subsubsection{Reactive layer}

As its name suggests, a variety of reactive learning processes take place on the reactive layer. The core idea of reaction learning within the MLOC architecture is to respond to external stimuli by using model-less reinforcement learning [11]. This complements the algorithms on the active layer, which rely on a relatively accurate model to exist in order to be effective. The anytime learning architecture on the contrary, is completely based around a simulation model and using any algorithms which do not utilize this would defy the purpose of anytime learning. For this reason, anytime learning has no equivalent components in particular to take care of those tasks. Nevertheless, it is possible to find some concepts from the reactive layer, such as the support for reinforcement learning, which anytime learning was designed for [3], within the architecture of anytime learning. As a whole, it might be most reasonable to establish that the reflective and, albeit to a much lesser extent, the reactive layer were merged into a single one in the case of anytime learning.

\subsubsection{Reflective layer} 
Finally, there is the \textit{reflective layer}. It consists of \textit{Learning} and \textit{Optimization} mechanisms. Both of these mechanism access the same models in order to improve the models based on their findings and as a consequence, improve the behavior of the system by learning from those models. Much like on the active layer, a lot of similarities to anytime learning can be found in this. Just like the MLOC architecture, anytime learning uses a simulation model that is continuously being adjusted in order to improve the performance of an agent. The \textit{simulation model}, \textit{decision maker model}, \textit{test knowledge} base and \textit{learning method} from Fig. 3 as a whole could be considered to form the equivalent of the reflective layer in anytime learning. 

\subsubsection{Neighborhood}
In addition to these layers which all belong to a single agent, the MLOC architecture includes a neighborhood consisting of other agents. anytime learning does not account for any neighborhood, nor does it communicate with other agents in any way, which is why the comparison between anytime learning and MLOC in this paper is limited to a single agent in this neighborhood.

All in all, anytime learning bears a strong resemblance to MLOC. While it is slightly more advanced than the basic observer/controller pattern, it lacks the reactive layer which is found in MLOC. Both architectures are based on splitting up the system into a learning observer and an optimizing controller, both of which are accessing the same (simulation) model. As such, one could consider anytime learning to be a slightly less sophisticated and earlier variant of MLOC.
Although some components which are included in MLOC are amiss in anytime learning, it seems evident that anytime learning nonetheless maps very closely to the MLOC architecture. From this can be concluded that, while modern architectures for organic computing are obviously more advanced than the architecture of anytime learning, the concept of anytime learning is a significant part of organic computing even today, and the principles found in it hold meaning still, even years after the approach was first suggested. Not only is anytime learning quite clearly an early approach to organic computing, but it still finds application nowadays.

\begin{figure}[htbp]
\centerline{\includegraphics[width=3.5in]{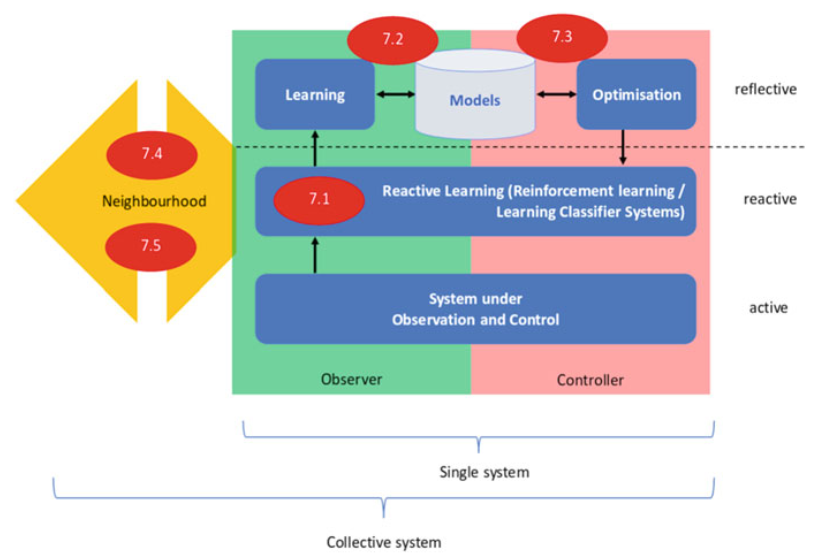}}
\caption{The MLOC model as it is introduced and illustrated in [11] bears a number of similarities to the anytime learning architecture.}
\label{fig}
\end{figure}

\section{Conclusion}
Is anytime learning the next step in organic computing? Although the roots of anytime learning reach far back and the concept was first presented more than two decades ago, many aspects of it still remain to be researched today, and the approach is not perfect just yet. Nevertheless, what has been shown already is the fact that anytime learning is able to deal with a large variety of issues which previously proved to be a challenge, if the concept is utilized correctly. The many variations of anytime learning which were based on Grefenstette's and Ramsey's work show how much potential the concept provides, but at the same time they highlight the uncertainties on how exactly the architecture should be implemented and how to deal with its weak points.

Anytime learning exhibits the properties that are required for organic computing to a high extent and lives up to the vision of IBM by doing so. More than that: It lives up to the vision of IBM long before this vision was even established. Therefore, while it can not be said for sure whether anytime learning will be the \textit{next} step in organic computing, it was most certainly one of the \textit{first} steps and many of its principles still find application today. What can be said for sure is that it provides lots of grounds for further research. If the uncertainties and problematic parts of the concept are addressed correctly, it might indeed turn out to not only be one of the first steps, but a concept that might find application in parts or as a whole in many more modern architectures, making it perhaps not \textit{the} next step, but one of the next steps in the grand scheme of organic computing.

\vspace{12pt}


\begin{thebibliography}{00}
\bibitem{b1} J. O. Kephart and D. M. Chess, "The vision of autonomic computing" in \textit{Computer}, vol. 36, no. 1, pp. 41-50, 2003
\bibitem{b2} A. G. Ganek and T. A. Corbi, "The dawning of the autonomic computing era" in \textit{IBM Systems Journal}, vol. 42, no. 1, pp. 5-18, 2003. 
\bibitem{b3} J.J. Grefenstette, C.L. Ramsey, "An Approach to Anytime Learning" in \textit{Proceedings of the Ninth International Conference on Machine Learning}, San Mateo, CA: Morgan Kaufmann, 1992, pp. 189–195
\bibitem{b4} S. Zilberstein, "Using Anytime Algorithms in Intelligent Systems", \textit{AI Magazine} vol. 17, no. 3, pp. 73-83, FALL 1996
\bibitem{b5} G. B. Parker, "Punctuated anytime learning for hexapod gait generation" in \textit{IEEE/RSJ International Conference on Intelligent Robots and Systems}, 2002, pp. 2664-2671 vol.3 
\bibitem{b6} J.J. Grefenstette, H.G. Cobb, \textit{User's Guide for SAMUEL, Version 1.3}, 1991
\bibitem{b7} C. L. Ramsey, J.J. Grefenstette, "Case-Based Anytime Learning" in \textit{AAAI Technical Report WS-94-01}, 1994, pp. 91-95
\bibitem{b8} A. Bonarini, "Anytime Learning and Adaption of Structured Fuzzy Behaviors" in \textit{Adaptive Behaviour } vol. 5, no. 1, pp. 281-315, 1997
\bibitem{b9} J. Branke \textit{et al.}, "Organic Computing – Addressing Complexity by Controlled Self-Organization" in \textit{Second International Symposium on Leveraging Applications of Formal Methods, Verification and Validation}, Paphos, 2006, pp. 185-191. 
\bibitem{b10} C. Mueller-Schloer "Organic computing: on the feasibility of controlled emergence" in: \textit{Proceedings of the 2nd IEEE/ACM/IFIP international conference on Hardware/software codesign and system synthesis}, ACM, 2004, pp. 2-5
\bibitem{b11} C. Mueller-Schloer, S. Tomforde, "Basic Methods" in \textit{Organic Computing – Technical Systems for Survival in the Real World}, Cham: Springer International Publishing AG, 2017
\end{thebibliography}
\end{document}